\documentclass[12pt]{iopart}
\expandafter\let\csname equation*\endcsname\relax
\expandafter\let\csname endequation*\endcsname\relax
\usepackage{amsmath}
\usepackage{bm}
\usepackage{iopams}

\usepackage{hyperref}
\usepackage[usenames,dvipsnames]{xcolor}
\hypersetup{colorlinks=true,citecolor=MidnightBlue, linkcolor=MidnightBlue, urlcolor=MidnightBlue} 
\usepackage{setspace}
\usepackage{graphicx}
\usepackage{doi}
\def\d{{\mathrm{d}}}

\makeatletter
\numberwithin{equation}{section}
\begin{document}
%------------------------------------------------------------------------------------------------------------------------------------------

\title[Cosmographic analysis of redshift drift]{
{\LARGE Cosmographic analysis of redshift drift}
}

\author{
\Large Francisco S. N. Lobo$\,^{1}$, 
Jos\'e Pedro Mimoso$\,^{2}$\\
{\sf  and} Matt Visser$\,^{3}$}

\address{
$^{1}$  
Instituto de  Astrof\'{i}sica e Ci\^encias do Espa\c co, \break
\null\qquad Faculdade de Ci\^{e}ncias da Universidade de Lisboa, Campo Grande, Ed. C8 \break
\null\qquad 1749-016 Lisboa, Portugal
}
\address{
$^{2}$ 
Departamento de F\'{i}sica and Instituto de  Astrof\'{i}sica e Ci\^encias do Espa\c co,\break
\null\qquad Faculdade de Ci\^{e}ncias da Universidade de Lisboa, Campo Grande, Ed. C8 \break
\null\qquad 1749-016 Lisboa, Portugal
}
\address{
$^{3}$ 
School of Mathematics and Statistics, Victoria University of Wellington, \break
\null\qquad PO Box 600, Wellington 6140, New Zealand
}

\ead{fslobo@fc.ul.pt;~jpmimoso@fc.ul.pt;~
%\\[0pt]\ \ \ \ \ \ \
matt.visser@sms.vuw.ac.nz}
%------------------------------------------------------------------------------------------------------------------------------------------
%------------------------------------------------------------------------------------------------------------------------------------------
%------------------------------------------------------------------------------------------------------------------------------------------
%------------------------------------------------------------------------------------------------------------------------------------------%------------------------------------------------------------------------------------------------------------------------------------------
%------------------------------------------------------------------------------------------------------------------------------------------
\begin{abstract}
Redshift drift is the phenomenon whereby the observed redshift between an emitter and observer comoving with the Hubble flow in an expanding FLRW universe will slowly evolve --- on a timescale comparable to the Hubble time. There are nevertheless serious astrometric proposals for actually observing this effect. We shall however pursue a more abstract theoretical goal, and perform a general cosmographic analysis of this effect, eschewing (for now) dynamical considerations in favour of purely kinematic symmetry considerations and Taylor series expansions based on FLRW spacetimes. We shall develop various exact results and series expansions for the redshift drift (and its derivatives) in terms of the present day Hubble, deceleration, jerk, snap, crackle, and pop parameters, as well as the present day redshift of the source. 
In particular, potential observation of this redshift drift effect is intimately related to the universe exhibiting a nonzero deceleration parameter.


\medskip
\noindent
Published as: JCAP \textbf{04} (2020), 043. 
\doi{10.1088/1475-7516/2020/04/043}

\medskip
\noindent
D{\sc{ate}}:  9 February 2020;  26 March 2020; \LaTeX-ed \today.

\end{abstract}

%------------------------------------------------------------------------------------------------------------------------------------------
%------------------------------------------------------------------------------------------------------------------------------------------
%------------------------------------------------------------------------------------------------------------------------------------------
\singlespacing
%------------------------------------------------------------------------------------------------------------------------------------------
\maketitle
%------------------------------------------------------------------------------------------------------------------------------------------
\tableofcontents
%------------------------------------------------------------------------------------------------------------------------------------------
\markboth{Cosmographic analysis of redshift drift\hfill}{}

\clearpage
%------------------------------------------------------------------------------------------------------------------------------------------
\def\d{{\mathrm{d}}}
%-----------------------------------------------------------------------------------------------------
\section{Introduction}
%-----------------------------------------------------------------------------------------------------
%---------------------------------------------------------------------------------------------------
\label{S:introduction}
%----------------------------------------------------------------------------------------------------
The concept of ``redshift drift'' dates back (at least) some 58 years, to 1962, arising in coupled papers by  Sandage~\cite{Sandage:1962} and McVittie~\cite{McVittie:1962}. 
Relatively little direct follow-up work took place in the 20$^{th}$ century, with Loeb's 1998 article~\cite{Loeb:1998} as a stand-out exception. However, over the last 15 years the concept of redshift drift has become much more mainstream~\cite{Linder:2008,Quercellini:2010,Uzan:2007,Uzan:2008,
Neben:2012,Alves:2019,Bolejko:2019,Liske:2008,Yoo:2010,Steinmetz:2008,Kim:2014,Killedar:2009,Koksbang:2015,Marcori:2018,Martins:2016,Koksbang:2019,Zhang:2016,Capozziello:2013}. 
The basic idea is this: If in a FLRW universe emitter and observer are comoving with the Hubble flow then the null curve connecting them is slowly evolving on a timescale set by the Hubble parameter ---
this implies that the redshift is slowly evolving. 
Despite the fact that the magnitude of the redshift drift is extremely small --- the spectral shift is of order one part in $10^{9}$ to $10^{10}$ over the period of a decade --- the  realistic possibility of the detectability of this effect has been explored in the subsequent literature~\cite{Loeb:1998,Linder:2008,Quercellini:2010,Uzan:2007,Uzan:2008,Neben:2012,Alves:2019,Bolejko:2019}.

The key equation (which we shall re-derive \emph{and subsequently extend} below) is:
\begin{equation}
\dot z  = (1+z) H_0 - H(z).
\end{equation}
See specifically McVittie, and in fact all of references~\cite{Sandage:1962,McVittie:1962,Loeb:1998,
Linder:2008,Quercellini:2010,Uzan:2007,Uzan:2008,
Neben:2012,Alves:2019,Bolejko:2019,Liske:2008,Yoo:2010,Steinmetz:2008,Kim:2014,Killedar:2009,Koksbang:2015,Marcori:2018,Martins:2016,Koksbang:2019}. 
A second key result at low redshift is presented by Neben \&\ Turner~\cite{Neben:2012}, citing McVittie~\cite{McVittie:1962}, where they assert
\begin{equation}
\dot z = - z \,q_0 \, H_0 + O(z^2).
\end{equation}
(See also Martins~\emph{et al.}~\cite{Martins:2016}.)
Note that this is intimately related to the deceleration parameter, so that the presence of a redshift drift is a direct signature of acceleration or deceleration.
We shall soon see that higher derivatives in the redshift drift are related to higher orders in the cosmographic expansion.

%  'jerk' (or sometimes, boringly, jolt, surge, or lurch) and is measured in units of m/s³. 
%
% Snap = ??spasm.?? It has also been called a ??jounce,?? a ??sprite,?? a ??surge,?? 
%or a ??snap,?? with its successive derivatives, ??crackle?? and ??pop.?? [17]
 %
% snap, crackle, pop, lock and drop...
% Shot (9th) 
%Put (10th)

Specifically we shall extend this result to include the jerk, snap, crackle, and pop; in principle we could go to even higher order in the cosmographic expansion. It should be noted that the terminology is not entirely uniform or standardized. 
Instead of using ``jerk'' to denote the third derivative $\dddot a$ some authors use ``jolt'' or ``super-acceleration'' or more rarely (and with significant risk of confusion) ``lurch'', ``pulse'', ``impulse'', ``bounce'', ``surge'', or ``shock''.  Instead of using ``snap'' to denote the fourth derivative $\ddddot a$ some authors use ``lerk'' or ``jounce''.  Instead of ``crackle'' for the fifth derivative $d^5a/dt^5$ some authors simply use ``$m$''.   There seems to be little to no competition for using ``pop'' to denote the sixth derivative $d^6a/dt^6$. 
Rather rarely, we have encountered specialized names for even higher derivatives: 
``lock'' ($d^7a/dt^7$), ``drop'' ($d^8a/dt^8$), ``shot'' ($d^9a/dt^9$), 
and ``put'' ($d^{10}a/dt^{10}$).
A discussion of some these terminological issues can be found in references~\cite{jerk1,jerk2,jerk3}. Our choice of terminology, jerk, snap, crackle, and pop, seems to be reasonably well-established in the wider community.

\clearpage

The only reason that we stop our expansions  at the sixth derivative ``pop'' is a purely pragmatic one --- the expansion formulae simply get too long and unwieldy.  In counterpoint, one does want to go to high enough order that one has a good basis for looking for possible patterns in the expansion.

One way of writing one of our central results is this:
\begin{eqnarray}
\fl \dot z &=& - H_0 z \left\{q_0 + {1\over2!}(j_0-q_0^2)z - {1\over3!}(s_0+3j_0+4j_0q_0-3q_0^2-3q_0^3) z^2   \right.
\nonumber
\\
\fl&&
+{1\over 4!}\{(c_0+8s_0+12j_0-4j_0^2)+(7s_0+32j_0)q_0 +(25j_0^2-12)q_0^2 -24q_0^3-15q_0^4 \}z^3
\nonumber
\\
\fl&&
 -{1\over 5!} \{(p_0+15c_0+60s_0+60j_0-15s_0j_0-60j_0^2)+(11c_0+105s_0+240j_0-70j_0^2)q_0
 \nonumber
\\
\fl&&
\qquad\qquad
 +(60s_0+375j_0-60)q_0^2+(210j_0-180)q_0^3-225q_0^4-105q_0^5\}z^4
  \nonumber
\\
\fl&&
\left.\qquad\qquad\vphantom{1\over1}
 + O(z^5) \right\}. \;\;
\end{eqnarray}
This determines the (first-order) redshift drift in terms of the present day Hubble, deceleration, jerk,  snap, crackle, and pop parameters, as well as the present day redshift of the source.

%\clearpage
The article is  outlined as follows: In Section~\ref{S:cosmoexp} we systematically develop 
cosmographic expansions for the emission time as a function of redshift $t(z)$, the Hubble  parameter $H(z)$, and for the deceleration, jerk, snap, crackle, and pop parameters.
Section~\ref{S:drift} presents various series expansions for the redshift drift in terms of the present day Hubble, deceleration, jerk, snap, crackle, and pop parameters, as well as the present day redshift of the source.  We discuss convergence criteria in Section \ref{S:convergence}, and introduce a modified $y$-redshift in Section~\ref{S:y-redshift}. Finally, we conclude in Section \ref{S:Conclusions}.

%----------------------------------------------------------------------------------------------------
\section{Cosmographic expansion}
\label{S:cosmoexp}
%----------------------------------------------------------------------------------------------------

%----------------------------------------------------------------------------------------------------
\subsection{Emission time as a function of redshift $t(z)$}\label{S:lookback}
%----------------------------------------------------------------------------------------------------

The idea of cosmography (cosmokinetics) dates back (at least) to Weinberg's 1972 textbook~\cite{Weinberg:1972}. The central idea is to maximize the use of the symmetries of FLRW spacetime,
\begin{equation}
ds^2 = - dt^2 + a(t)^2 
\left\{ {dr^2\over1-kr^2} +r^2 (d\theta^2+\sin^2\theta \; d\phi^2) \right\},
\end{equation}
 and delay the explicit use of the Einstein equations for as long as possible. Cosmographic ideas have become increasingly popular over the last two decades. See references~\cite{jerk1,jerk2} and~\cite{Blandford:2004,Visser:2004,Aviles:2012,Vitagliano:2009,Cattoen:2007a,Cattoen:2007b,Cattoen:2008,Visser:2009,Capozziello:2008,Dunsby:2015,Xia:2011,Vitagliano:2013,Capozziello:2017,Capozziello:2019}.
 
% \enlargethispage{10pt}
\clearpage
 Our first goal will be to invert the standard relationship $1+z = {a_0\over a(t)}$ to find $t(z)$  which we can formally define as $t(z) = a^{-1}(a_0/(1+z))$. We shall aim for a power series expansion of $t(z)$. 
Using only the symmetries of FLRW spacetime, together with the \emph{definitions}
\begin{eqnarray}
&& H = {\dot a\over a}; \qquad \qquad \qquad\;\;
q = -{a\; \ddot a\over (\dot a)^2}= -{\ddot a\over a\,H^2}; \qquad 
j = {a^2 \; \dddot a\over (\dot a)^3}= {\dddot a\over a\,H^3}; 
\nonumber\\
&&
s = {a^3 \;\ddddot a\over (\dot a)^4} = {\ddddot a\over a\,H^4}; \qquad
c= {a^4 \;  a^{(5)}\over (\dot a)^5}= { a^{(5)}\over a\,H^5}; \qquad
p= {a^5 \;  a^{(6)}\over (\dot a)^6}= { a^{(6)}\over a\,H^6}; \qquad
\end{eqnarray}
of the Hubble, deceleration, jerk, snap, crackle, and pop parameters, 
a (truncated)  Taylor series expansion around the current epoch yields
\begin{eqnarray}
\fl a(t) &=& a_0 \left\{\vphantom{1\over1} 1 +H_0(t-t_0) - {q_0\over2!} [H_0(t-t_0)]^2 + {j_0\over3!} [H_0(t-t_0)]^3
+{s_0\over4!} [H_0(t-t_0)]^4\right.
\nonumber
\\
\fl&& \qquad\qquad\left.\vphantom{1\over1}
+{c_0\over5!} [H_0(t-t_0)]^5+{p_0\over6!} [H_0(t-t_0)]^6+ O([t-t_0]^7) \right\}. 
\end{eqnarray}

Note that the only reason for the presence of the minus sign in the definition of the \emph{deceleration} parameter $q$ is a purely historical one; current belief is that asymptotically the deceleration parameter will become negative, so that it would make sense to define $\alpha=-q=+{\ddot a a \over \dot a^2}$ as an \emph{acceleration} parameter. This would eliminate a few annoying minus signs in the expansions below, but such a terminological shift would be sufficiently drastic to cause significant confusion.
So we shall keep the usual definition of the deceleration parameter $q$, with its inherent minus sign.

Relating this Taylor expansion of $a(t)$ to the redshift in the usual way, $a/a_0=1/(1+z)$,  we have the utterly  standard result
\begin{eqnarray}
\fl{1\over1+z} &=& 1 +H_0(t-t_0) - {q_0\over2!} [H_0(t-t_0)]^2 + {j_0\over3!} [H_0(t-t_0)]^3
+{s_0\over4!} [H_0(t-t_0)]^4
\nonumber
\\
\fl&& \qquad\qquad +{c_0\over5!} [H_0(t-t_0)]^5+{p_0\over6!} [H_0(t-t_0)]^6+ O([t-t_0]^7). 
\end{eqnarray}
Note the presence, on the left hand side of the equation,  of a pole at $z=-1$, which corresponds to $a\rightarrow \infty$, that is, the instant that the universe has expanded to  infinite size.

\bigskip
This series is now easily \emph{reverted}~\cite{reversion1,reversion2}  to yield emission time as a function of redshift. 

To order $O(z^6)$, that is, including terms up to the pop $p_0$,  one has:
\begin{eqnarray}
\fl t(z) &=& t_0 + {z\over H_0} \left\{-1+\left(1+{q_0\over2!}\right)z -\left(1-{j_0-6q_0-3q_0^2\over3!}\right)z^2 
\right.
\nonumber\\
\fl &&
+
\left(1 - {s_0 +12j_0 -36q_0 +10j_0q_0 -36q_0^2 -15q_0^3\over4!}\right)z^3 
\nonumber\\
\fl &&
-
\left(1 - {1\over5!}\{(c_0 +20s_0 +120j_0- 10j_0^2) +(15s_0+200j_0-240)q_0\right.
\nonumber\\
\fl &&
\qquad\qquad
\left.\vphantom{1\over1}
  +(105j_0-360)q_0^2 -300q_0^3 -105q_0^4\}\right)z^4 \
\nonumber\\
\fl &&
+
\left(1 - {1\over6!} \{ (p_0+30c_0+300s_0+1200j_0-35j_0s_0-300j_0^2)
\right.
\nonumber\\
\fl &&
\qquad\qquad
+(21c_0+450s_0+3000j_0-1800-280j_0^2) q_0 
\nonumber\\
\fl &&
\qquad\qquad
+(3150j_0+210s_0 -3600)q_0^2 
\nonumber\\
\fl &&
\qquad\qquad
\left.\left.\vphantom{1\over1}
+(1260j_0-4500)q_0^3 -3150q_0^4-945q_0^5
\}\right)z^5 + O(z^6)  \right\}.
\end{eqnarray}
While this reversion could in principle be done by hand, at least for the first few terms,
use of a symbolic algebra package is certainly advantageous.
In contrast to what happens for luminosity distance, this expansion for $t(z)$ does not depend on the spatial curvature parameter $k\in\{-1,0,+1\}$.
Note that the quantity $t_0-t(z)$ is often called the ``lookback time''. 

We have gone to such a high order in the cosmographic expansion largely in the hope of finding useful patterns in the coefficients. 
One immediately useful pattern is the alternating $\pm 1$ leading terms at each order in redshift. 
However it must be admitted that the series expansion of $t(z)$ in terms of $z$ quickly becomes somewhat clumsy. 
%%%%%%%

%\clearpage
%\null
%\vspace{-25pt}

It is useful to note that the series for $t(z)$ can be partially summed as follows:
\begin{eqnarray}
\fl t(z) &=& t_0  - {1\over H_0}\; { z\over 1+z}    
+ {z^2\over H_0} \left\{{q_0\over2!} +\left({j_0-6q_0-3q_0^2\over3!}\right)z 
\right.
\nonumber\\
\fl &&
-
\left( {s_0 +12j_0 -36q_0 +10j_0q_0 -36q_0^2 -15q_0^3\over4!}\right)z^2
\nonumber\\
\fl &&
+
\left({1\over5!}\{(c_0 +20s_0 +120j_0- 10j_0^2) +(15s_0+200j_0-240)q_0\right.
\nonumber\\
\fl &&
\qquad\qquad
\left.\vphantom{1\over1}
  +(105j_0-360)q_0^2 -300q_0^3 -105q_0^4\}\right)z^3 \
\nonumber\\
\fl &&
-
\left({1\over6!} \{ (p_0+30c_0+300s_0+1200j_0-35j_0s_0-300j_0^2)
\right.
\nonumber\\
\fl &&
\qquad\qquad
+(21c_0+450s_0+3000j_0-1800-280j_0^2) q_0 
\nonumber\\
\fl &&
\qquad\qquad
+(3150j_0+210s_0 -3600)q_0^2 
\nonumber\\
\fl &&
\qquad\qquad
\left.\left.\vphantom{1\over1}
+(1260j_0-4500)q_0^3 -3150q_0^4-945q_0^5
\}\right)z^4 + O(z^5)  \right\}.
\end{eqnarray}
This is perhaps the first indication that the variable $y={z\over1+z}$ may prove useful. 

A distinct framework that is plausibly cosmographic in nature is the use of  Pad\'e rational polynomial approximants, both  for the Hubble function $H(z)$ itself,  and also for other related cosmological functions such as $q(z)$. See references~\cite{Gruber:2013,Aviles:2014,Capozziello:2018,Benetti:2019}. 
A drawback of the Pad\'e approximant approach is that the coefficients in the Pad\'e rational polynomials typically do not have an immediate physical interpretation. 
So in the present article we shall work with the usual implementation of cosmography in terms of Taylor series.

%----------------------------------------------------------------------------------------------------
\subsection{Hubble  parameter $H(z)$}
\label{S:Hubble}
%----------------------------------------------------------------------------------------------------

Inserting the truncated Taylor series for $t(z)$ into the definition of the Hubble parameter $H(t)= \dot a(t)/a(t)$ and expanding one finds
\begin{eqnarray}
\fl H(z) &=& H_0 \left\{\vphantom{1\over1}1 + (1+q_0)z + {1\over2!}(j_0-q_0^2) z^2 
- {1\over3!} \left(s_0 +3 j_0 + 4j_0q_0 -  3 q_0^2  - 3 q_0^3 \right) z^3 
\right.
\nonumber\\
\fl &&
\qquad \vphantom{1\over1} 
+{1\over4!} \{(c_0+8s_0+12j_0-4j_0^2)+(7s_0+32j_0)q_0 +(25j_0-12) q_0^2 
\nonumber\\
\fl &&
\qquad \qquad\qquad\vphantom{1\over1} 
-24 q_0^3 -15q_0^4\} z^4 
\nonumber\\
\fl &&
\qquad \vphantom{1\over1} 
-{1\over5!} \{(p_0+15c_0+60s_0+ 60 j_0 - 60 j_0^2 -15 j_0 s_0)  
\nonumber\\
\fl &&
\qquad\qquad\qquad \vphantom{1\over1} 
+(11c_0+105s_0+240j_0-70j_0^2)q_0
\nonumber\\
\fl &&
\qquad \qquad\qquad\vphantom{1\over1} 
+(60s_0+375j_0-60)q_0^2  +(210j_0-180)q_0^3
\nonumber\\
\fl &&
\qquad\qquad\qquad \left.\vphantom{1\over1} 
 -225 q_0^4 -105 q_0^5\} z^5 + O(z^6) \right\}.
\end{eqnarray}

Note that this result for $H(z)$ again does not depend on the spatial curvature parameter $k\in\{-1,0,+1\}$.
This expansion is purely cosmographic, no dynamics is required in deriving this result.
The expansion for $H(z)$ can easily be extended to higher order in the redshift, it just becomes increasingly more  tedious and messy to write down.

%----------------------------------------------------------------------------------------------------
\subsection{Deceleration, jerk, snap, crackle, and pop parameters in terms of $z$}
\label{S:deceleration-to-pop}
%----------------------------------------------------------------------------------------------------

In a similar fashion one easily derives  cosmographic expansions for the
deceleration, jerk, snap, crackle, and pop parameters.

\paragraph{Deceleration:}
\begin{eqnarray}
\fl q(z) &=& q_0 +(j_0-q_0-2 q_0^2) z - 
{1\over2!}\left(s_0+4j_0-2q_0 +{7} q_0j_0-8q_0^2-8 q_0^3 \right) z^2 
\nonumber\\
\fl&&
+{1\over 3!}  \big[(c_0+9s_0+18j_0-7j_0^2) +(11s_0+63j_0-6)q_0 + (59j_0-36) q_0^2 
\nonumber\\
\fl&&
\qquad\qquad -72 q_0^3 - 48q_0^4 \big] z^3 
\nonumber\\
\fl&&
- {1\over 4!}\big[(p_0+16c_0+72s_0+96j_0-25s_0j_0-112j_0^2)
\nonumber\\
\fl&&
\qquad\qquad+(16c_0+176s_0+504j_0-160j_0^2-24)q_0
\nonumber\\
\fl&&
\qquad\qquad+(125s_0+944 j_0-192)q_0^2 + (605j_0-576)q_0^3 -768 q_0^4- 384 q_0^5 \big] z^4
\nonumber\\
\fl&&
+ O(z^5).
\end{eqnarray}

\paragraph{Jerk:}
\begin{eqnarray}
\fl j(z) &=& j_0 -(s_0+2j_0+3 q_0j_0) z 
+  \frac{1}{2!}  \big[(c_0+6s_0+6j_0-3j_0^2) 
%\nonumber \\
%&&
%\qquad 
+(7s_0+18j_0)q_0 +15 j_0 q_0^2 \big] z^2
\nonumber\\
\fl&& - \frac{1}{3!} \big[(p_0+12c_0+36s_0+24j_0-13s_0j_0-36j_0^2)
+(12c_0+108j_0+84s_0-48j_0^2 )q_0 
\nonumber\\
\fl&&
\qquad +(57s_0+180j_0)q_0^2 +105 j_0 q_0^3 \big] z^3 + O(z^4).
\end{eqnarray}

\paragraph{Snap:}
\begin{eqnarray}
s(z) &=& s_0 - (c_0+3s_0+4s_0q_0)z + \frac{1}{2!}  \big[ (p_0+8c_0+12s_0-4s_0j_0)
\nonumber\\
&&+(9c_0+32s_0)q_0+24s_0q_0^2 \big]z^2 +  O(z^3).
\end{eqnarray}

\paragraph{Crackle:}
\begin{equation}
c(z) = c_0 - (p_0+4c_0+5 c_0q_0)z + O(z^2).
\end{equation}

\paragraph{Pop:}
\begin{equation}
p(z) = p_0 + O(z).
\end{equation}
Despite the relatively messy form of some of these expansions, there are some definite patterns here. 

%\clearpage
For instance, working at lowest non-trivial order in redshift, from the above we see
\begin{eqnarray}
q(z) &=&q_0 -{\dot q_0\over H_0}\; z + O(z^2);
\\
j(z) &=& j_0 -{d j_0/dt\over H_0}\; z + O(z^2);
\\
s(z) &=&s_0 -{\dot s_0\over H_0}\; z + O(z^2);
\\
c(z) &=&c_0 -{\dot c_0\over H_0}\; z + O(z^2).
\end{eqnarray}
In fact with a bit more work
\begin{equation}
p(z) =p_0 -{\dot p_0\over H_0}\; z + O(z^2).
\end{equation}
This can easily be generalized to $n^{th}$ order.

%----------------------------------------------------------------------------------------------------
\section{Redshift drift in terms of $z$}\label{S:drift}
%----------------------------------------------------------------------------------------------------
To see where the redshift drift comes from, start with the utterly standard FLRW result
\begin{equation}
1+z = {a_{0}\over a_{e}} = {dt_{0}\over dt_{e}}.
\end{equation}
Here the subscript 0 denotes the current epoch (reception of the photon) while the subscript $e$ denotes the emission event.

%----------------------------------------------------------------------------------------------------
\subsection{First-order redshift drift}
%----------------------------------------------------------------------------------------------------
By the chain rule we have
\begin{equation}
\dot z = {\dot a_{0}\over a_{e}} - {a_{0} (d a_{e}/dt_{0})\over a_{e}^2}
={\dot a_{0}\over a_{e}} - {a_{0} (d a_{e}/dt_{e}) (dt_{e}/dt_{0})\over a_{e}^2}.
\end{equation}
Simplifying 
\begin{equation}
\dot z = {\dot a_{0}\over a_{0}} \; {a_{0}\over a_{e}}
 -  {da_{e}/dt_{e} \over a_{e}}
 = (1+z) \;H_0 - H_{e}.
\end{equation}
That is
\begin{equation}
\dot z  = (1+z) \;H_0 - H(z).
\end{equation}
This is McVittie's result~\cite{McVittie:1962}. 

%\clearpage
Once we have this key exact result, combining this with our cosmographic expansion for $H(z)$ easily yields
\begin{eqnarray}
\fl \dot z &=& - H_0 z \left\{q_0 + {1\over2!}(j_0-q_0^2)z - {1\over3!}(s_0+3j_0+4j_0q_0-3q_0^2-3q_0^3) z^2   \right.
\nonumber
\\
\fl&&
+{1\over 4!}\left[(c_0+8s_0+12j_0-4j_0^2)+(7s_0+32j_0)q_0 +(25j_0-12)q_0^2 -24q_0^3-15q_0^4 \right] z^3
\nonumber
\\
\fl&&
 -{1\over 5!} \big[(p_0+15c_0+60s_0+60j_0-15s_0j_0-60j_0^2)+(11c_0+105s_0+240j_0-70j_0^2)q_0
 \nonumber
\\
\fl&&
\qquad\qquad
 +(60s_0+375j_0-60)q_0^2+(210j_0-180)q_0^3-225q_0^4-105q_0^5\big] z^4
  \nonumber
\\
\fl&&
\left.\qquad\qquad
 + O(z^5) \vphantom{1\over1} \right\}. \;\;
\end{eqnarray}
The lowest-order term is the Neben \&\ Turner~\cite{Neben:2012} result
\begin{equation}
\dot z = - z \,q_0 \, H_0 + O(z^2).
\end{equation}
Note that the timescale for the redshift drift is of order the Hubble time.
This makes potential observations challenging~\cite{Linder:2008,Uzan:2007},
though other authors are considerably more optimistic~\cite{Loeb:1998,Quercellini:2010,Neben:2012,Alves:2019}. 

%----------------------------------------------------------------------------------------------------
\subsection{Second-order redshift drift}
%----------------------------------------------------------------------------------------------------
We can evaluate $\ddot z$ as follows:
\begin{equation}
\ddot z = {d\over dt}\left[ (1+z) H_0 - H(z) \right] = \dot z H_0 + (1+z) \dot H_0 - \dot H(z).
\end{equation}
In the usual manner
\begin{equation}
\dot H_0 = {d\over dt}\left(\dot a\over a\right) = -  (1+q_0)H_0^2. 
\end{equation}
A trifle more subtle is the chain rule result
\begin{equation}
\dot H(z) = {dt_{e}\over dt} \; {d\over dt_{e}}  \left(\dot a\over a\right)_{e} 
= - {1+q(z)\over1+z} H(z)^2. 
\end{equation}
Combining these results
\begin{eqnarray}
\ddot z &=& \dot z H_0 -(1+z)H_0^2(1+q_0) + {1+q(z) \over1+z} \; H(z)^2
\nonumber\\
&=& 
\left((1+z)H_0- H(z)\right) H_0 -(1+z)H_0^2(1+q_0) + {1+q(z) \over1+z} H(z)^2
\nonumber\\
&=& 
- H_0 H(z) - (1+z) q_0 H_0^2 + \left[1+q(z)\over1+z\right] H(z)^2.
\end{eqnarray}
That is
\begin{equation}
\ddot z =  - (1+z) q_0 H_0^2 - H_0 H(z) + \left[1+q(z)\over1+z\right] H(z)^2.
\end{equation}

This formula is, within the framework of FLRW spacetimes, exact.

Inserting our previous expansions for $H(z)$ and $q(z)$ one now obtains the cosmographic expansion
\begin{eqnarray}
\fl\ddot z &=& z \; H_0^2 \left\{j_0 -{1\over2!}(s_0+j_0+j_0q_0-q_0^2) z \right.
  \nonumber
\\
\fl&&
+{1\over 3!}((c_0+5s_0+3j_0-j_0^2)+(3s_0+8j_0)q_0+(3j_0-3)q_0^2 -3q_0^3   )z^2 
 \nonumber
\\
\fl&&
-{1\over4!}((p_0+11c_0+28s_0+12j_0-5s_0j_0-14j_0^2)
+(6c_0+37s_0+52j_0-10j_0^2)q_0
  \nonumber
\\
\fl&&
\left.\qquad\qquad\vphantom{1\over1}
+(15s_0+55j_0-12)q_0^2 
+ (15j_0-24)q_0^3 - 15 q_0^4)z^3+ O(z^4)  \right\}.
\end{eqnarray}

At lowest order this agrees with Martins~\emph{et al.}~\cite{Martins:2016}, who in their equation (24) state
\begin{equation}
\ddot z = z\,j_0 \, H_0^2 + O(z^2).
\end{equation}

%----------------------------------------------------------------------------------------------------
\subsection{Third-order redshift drift}
%----------------------------------------------------------------------------------------------------

Differentiating yet a third time we obtain
\begin{eqnarray}
\dddot z = {d\over dt}\left[  - (1+z) q_0 H_0^2 - H_0 H(z) + \left[1+q(z)\over1+z\right] H(z)^2 \right].
\end{eqnarray}
Now we have already seen
\begin{equation}
\dot H_0 = - (1+q_0)H_0^2; \qquad \dot H(z) = - {1+q(z)\over1+z} \; H(z)^2.
\end{equation}
The new ingredient is
\begin{equation}
\fl \dot q_0 = -(j_0-q_0-2q_0^2) H_0; \qquad \dot q(z) = -{1\over1+z} \;\left[j(z)-q(z)-2q(z)^2\right] H(z).
\end{equation}
Combining these results we see
\begin{eqnarray}
\dddot z &=& -\dot z q_0 H_0^2 + (1+z)(j_0-q_0-2q_0^2) H_0^3 +2(1+z)q_0(1+q_0)H_0^3
\nonumber
\\
&& \qquad 
+(1+q_0) H_0^2 H(z) + {1+q(z)\over 1+z} H_0 H(z)^2
- \dot z \left[1+q(z)\over(1+z)^2\right] H(z)^2
\nonumber\\
&& \qquad  - \left[j(z)-q(z)-2q(z)^2 \over(1+z)^2\right] H(z)^3 - 2\left[1+q(z)\over1+z\right]^2 H(z)^3.
%\nonumber
\end{eqnarray}
That is
\begin{eqnarray}
\fl \dddot z &=& - q_0 [(1+z)H_0-H(z)] H_0^2 + (1+z)(j_0-q_0-2q_0^2) H_0^3 +2(1+z)q_0(1+q_0)H_0^3
\nonumber
\\
\fl && \quad 
+(1+q_0) H_0^2 H(z) + {1+q(z)\over 1+z} H_0 H(z)^2
- \left[1+q(z)\over(1+z)^2\right] [(1+z)H_0-H(z)] H(z)^2
\nonumber\\
\fl && \qquad  - \left[j(z)-q(z)-2q(z)^2 \over(1+z)^2\right] H(z)^3 - 2\left[1+q(z)\over1+z\right]^2 H(z)^3.
%\nonumber
\end{eqnarray}
There are a significant number of cancellations, leading to the relatively pleasant result
\begin{equation}
\dddot z = (1+z)j_0H_0^3 + (1+2q_0)H_0^2 H(z)
- {\left[1+j(z) +2q(z)\right]H(z)^3\over(1+z)^2}.
\end{equation}
This formula is, within the framework of FLRW spacetimes, exact.

Inserting our previous expansions for $H(z)$, $q(z)$, and $j(z)$ one obtains
\begin{eqnarray}
\fl \dddot z &=& z \, H_0^3\left\{ s_0 -{1\over2!}((c_0+2s_0)+ (s_0+2j_0)q_0 ) z  
\right.
\nonumber\\
\fl&&
+{1\over3!} ((p_0+7c_0+8s_0-s_0j_0-4j_0^2)+(3c_0+9s_0+8j_0)q_0+(3s_0+6j_0)q_0^2) z^2
\nonumber\\
\fl&&
\left.
+ O(z^3)\right\} .
\end{eqnarray}

Note that at lowest order
\begin{equation}
\dddot z = z\,s_0\,H_0 + O(z^2). 
\end{equation}

%----------------------------------------------------------------------------------------------------
\subsection{Fourth-order redshift drift}
%----------------------------------------------------------------------------------------------------

We start by noting
\begin{equation}
\fl \ddddot z = {d\over dt} \left\{ (1+z)j_0H_0^3 + (1+2q_0)H_0^2 H(z)
- {\left[1+j(z) +2q(z)\right]H(z)^3\over(1+z)^2}\right\}.
\end{equation}
We already have
\begin{equation}
\dot H_0 = - (1+q_0)H_0^2; \qquad \dot H(z) = - {1+q(z)\over1+z} \; H(z)^2;
\end{equation}
and
\begin{equation}
\fl
\dot q_0 = -(j_0-q_0-2q_0^2) H_0; \qquad \dot q(z) = -{1\over1+z} \;\left[j(z)-q(z)-2q(z)^2\right] H(z).\qquad
\end{equation}
The new ingredient is
\begin{equation}
\fl
{d \,j_0\over dt} = (s_0+(2+3q_0)j_0) H_0; \qquad 
{ d\,j(z)\over dt} = {1\over1+z} \;\left[s(z) +(2+3q(z))j(z)\right] H(z).
\end{equation}
Thence, a little tedious algebra leads to
\begin{eqnarray}
\fl \ddddot z &=& (1+z)s_0H_0^4-(2+3j_0+4q_0)H_0^3H(z)
-{(1+q(z))(1+2q_0)\over1+z}H_0^2H(z)^2 
\nonumber\\
\fl&&
+{2(1+j(z)+2q(z))\over(1+z)^2} H_0 H(z)^3
\nonumber\\
\fl&&
+{1-s(z)+j(z)+3q(z)+2q(z)^2\over (1+z)^3} H(z)^4.
\end{eqnarray}
This result is still (within the framework of a FLRW universe) exact. 

Inserting our cosmographic series
\begin{equation}
\ddddot z = z \, H_0^4 \left\{ c_0  -{1\over2!} \left[(p_0 +3c_0-2 j_0^2)+(c_0 +s_0) q_0 \right]  z + O(z^2) \right\}.
\end{equation}
Note that at lowest order, as we have by now begun to expect,
\begin{equation}
\ddddot z = z \, c_0 \, H_0^4 + O(z^2).
\end{equation}

%----------------------------------------------------------------------------------------------------
\subsection{Fifth-order redshift drift}
%----------------------------------------------------------------------------------------------------
At fifth-order it is useful to simplify the argument by considering
\begin{equation}
z^{(5)} = {d\over dt} \left\{ \ddddot z\right\} 
= {d\over dt} \left\{  z \, c_0 \, H_0^4 + O(z^2)\right\}
= {d\over dt} \left\{  z \, c_0 \, H_0^4\right\} +O(z^2).
\end{equation}
Here we have used the fact that $\dot z = O(z)$.
Then
\begin{equation}
z^{(5)} 
= \left\{  \dot z \, c_0 \, H_0^4+ z \dot c_0 H_0^4 + 4 z c_0H_0^3 \dot H_0\right\} +O(z^2).
\end{equation}
But $\dot z = -z q_0 H_0+O(z^2)$ and $\dot H_0 = -(1+q_0)H_0^2$ while
\begin{equation}
\dot c_0 = \left[ p_0 + (4+5 q_0) c_0 \right] H_0.
\end{equation}
Combining the above
\begin{equation}
z^{(5)} = z \, p_0 \, H_0^5 + O(z^2).
\end{equation}
We shall now extend this to a general $n^{th}$-order result.

%----------------------------------------------------------------------------------------------------
\subsection{$n^{th}$-order redshift drift}
%----------------------------------------------------------------------------------------------------

Let us now define an $n^{th}$-order dimensionless generalized acceleration parameter, which we evaluate for convenience at the current epoch,  as
\begin{equation}
k_n = \left({a^{(n)}(t)\; a(t)^{n-1}\over [\dot a(t)]^n}\right)_{t=t_0}. 
\end{equation}
Here $a^{(n)}(t)$ denotes as usual the $n^{th}$ derivative.
Then $k_1=1$, $k_2=-q_0$, $k_3 = j_0$, and $k_4 = s_0$, $k_5=c_0$, and $k_6=p_0$.

Based on what we have already seen above, it seems plausible that the redshift drift satisfies 
\begin{equation}
z^{(n)} = z \;k_{n+1} \; H_0^n + O(z^2); \qquad \forall n \geq 1. 
\end{equation}
Certainly, as explicitly verified above, this is true for $n\in \{1,2,3,4,5\}$, and we shall now extend this to arbitrary $n$  by induction. First note that
\begin{equation}
\dot k_n =  \left({a^{(n+1)}\; a^{n-1}\over (\dot a)^n}\right)_0 
+ (n-1)  \left({a^{(n)}\; a^{n-2}\over (\dot a)^{n-1}}\right)_0 
- n  \left({a^{(n)}\; a^{n-1} \ddot a\over (\dot a)^{n+1}}\right)_0,
\end{equation}
which we can recast as
\begin{equation}
\dot k_n =  \left\{ k_{n+1} + (n-1) k_n + n q_0 k_n \right\} H_0. 
\end{equation}
But from the  discussion above we also know
\begin{equation}
\dot H_0 = - (1+q_0) H_0^2; \qquad \dot z = - z q_0 H_0 + O(z^2)
\end{equation}
So if we assume the induction hypotheses then
\begin{equation}
z^{(n+1)}(t) = {d z^{(n)}(t)\over dt} = {d\over dt}\left(z \;k_{n+1} \; H_0^n\right) + O(z^2)
\end{equation}
But then
\begin{eqnarray}
\fl {d\over dt}\left(z \;k_{n+1} \; H_0^n\right) &=& 
- z q_0 k_{n+1} H_0^{n+1} + z \left\{ k_{n+2} + n k_{n+1} + (n+1) q_0 k_{n+1} \right\} H_0^{n+1} 
\nonumber\\
\fl &&
- n z (1+q_0) k_{n+1} H_0^{n+1}  +O(z^2) 
\nonumber\\
\fl &=&
- z k_{n+2} H_0^{n+1} + O(z^2). 
\end{eqnarray}
This completes the proof of the inductive step. 

In view of the previous explicit verification for $n\in\{1,2,3,4,5\}$
this now completes the full proof that
\begin{equation}
z^{(n)} = z \;k_{n+1} \; H_0^n + O(z^2); \qquad \forall n \geq 1. 
\end{equation}
While direct measurement of these higher-order redshift drifts $z^{(n)}$ is likely to be technologically infeasible, they do have a nice theoretical interpretation in terms of the cosmographic parameters.

%----------------------------------------------------------------------------------------------------
\section{Convergence issues}\label{S:convergence}
%----------------------------------------------------------------------------------------------------

One of the problematic issues with cosmographic methods is that in the usual (naive) formulation one is dealing with truncated Taylor series in $z$ but often wishes to apply the formulae at large redshift $z>1$. Does the Taylor series converge? 
\emph{In fact, there are good mathematical and physical reasons for believing that these Taylor series in terms of $z$ cannot possibly converge for $z>1$.} See particularly references~\cite{Cattoen:2007a,Cattoen:2007b,Cattoen:2008}.
This follows from a variant of the Dyson argument~\cite{Dyson}  that is normally used in quantum field theory (QFT) to argue that the Feynman diagram expansion cannot possibly be convergent. Even after renormalization to eliminate the infinities, the Feynman diagram expansion is at best asymptotic.

In the present cosmographic context we argue as follows: If any of these Taylor series, (either for $t(z)$, $H(z)$, $q(z)$, $j(z)$, $s(z)$, $c(z)$, $p(z)$, $\dot z(z)$, or any of the $z^{(n)}$), were to converge for some region $z< z_*$ with $z_*>1$ then it is a standard result of real (or complex) analysis that the Taylor series must also converge for the reflected region $z>-z_*$ with $-z_*<-1$. But $z=-1$ corresponds to infinite expansion, so $z <-1$ corresponds to making predictions \emph{after} the universe has reached infinite size, which is physically unreasonable.

We can formulate this more precisely in terms of the radius of convergence $R_*$, which is determined by the distance from the origin $z=0$ to the nearest mathematical singularity. Looking into the future, suppose the universe has a future singularity, or turnaround event, or asymptotically approaches some finite size, at some $a_{\scriptscriptstyle \mathrm{max}}>a_0$, where we set $a_{\scriptscriptstyle \mathrm{max}}\to\infty$ if the universe expands to infinite size.
Then the Taylor series in $z$ converges for $|z|< R_*$ where we bound $R_*$ by
\begin{equation}
R_* = |z_{\scriptscriptstyle \mathrm{nearest\;singularity}}| 
\leq \left|{a_0\over a_{\scriptscriptstyle \mathrm{max}}}-1\right| 
= 1- {a_0\over a_{\scriptscriptstyle \mathrm{max}}} \leq 1.
\end{equation}
Since certainly $R_*\leq 1$, it makes no sense to push the Taylor series expansion into the region $z>1$. (In principle we should also look for past singularities, but since looking to the future already gives the convergence bound $|z|<1$, and since we have strong confidence in the \emph{absence} of physical singularities in the cosmologically recent past, such considerations are unnecessary for present purposes.)
Fortunately there are workarounds to side-step this convergence issue~\cite{Cattoen:2007a,Cattoen:2007b,Cattoen:2008}. 
Basically, one should rearrange the Taylor series to improve convergence. 
Indeed, mathematicians have developed an impressively large body of techniques for dealing with naively divergent series.
(See for instance~\cite{divergent-series}.) 

%----------------------------------------------------------------------------------------------------
\section{Modified $y$-redshift}\label{S:y-redshift}
%----------------------------------------------------------------------------------------------------

A physically well-motivated improved redshift parameterization that was mooted in references~\cite{Cattoen:2007a,Cattoen:2007b,Cattoen:2008} is to set
\begin{equation}
1-y = {a\over a_0} = {1\over 1+z},
\end{equation}
so that 
\begin{equation}
y = {z\over 1+z}; \qquad z = {y\over1-y}.
\end{equation}

Physically, in terms of the change in wavelength this corresponds to
\begin{equation}
z= {\Delta\lambda\over\lambda_e}; \qquad\qquad y = {\Delta\lambda\over\lambda_0}. 
\end{equation}
So when working with $y$ instead of $z$ all one is doing is redefining the redshift by normalizing it in terms of the arguably more physically relevant  present-day value of the wavelength~\cite{Cattoen:2007a,Cattoen:2007b,Cattoen:2008}.  Though physically equivalent to $z$, the $y$-redshift has much better mathematical convergence properties.

Now suppose the universe has a past singularity, or turnaround event, or asymptotically approaches some finite size, at some $a_{\scriptscriptstyle \mathrm{min}}<a_0$, where we set $a_{\scriptscriptstyle \mathrm{min}}\to0$ if the universe emerges from a big bang singularity.
Then the Taylor series in $y=z/(1+z)$ converges for $|y|< R_*$ where now
\begin{equation}
R_* = |y_{\scriptscriptstyle \mathrm{nearest\;singularity}}| 
\leq 
1- {a_{\scriptscriptstyle \mathrm{min}}\over a_0} \leq 1.
\end{equation}
But this corresponds to convergence, of the power series in $y$, for $z$ in the asymmetric region
\begin{equation}
z \in \left( - {R_*\over 1+R_*}, + {R_*\over 1-R_*}\right) \subseteq \left(-{1\over2}, \infty\right).
\end{equation}

Other functional relationships $z(y)$ and $y(z)$ have been mooted by other authors, see for instance references~\cite{Capozziello:2013} and~\cite{Aviles:2012}. In this article we concentrate on the particularly simple choice $z =y/(1-y)$, $y=z/(1+z)$ because it does minimal physical violence to the usual notion of redshift, being essentially just a change in normalization.
Furthermore with this definition of the $y$-redshift the mathematical analysis of convergence properties is particularly easy. 
Alternative choices mooted in the literature include
\begin{equation}
y(z) = \tan^{-1}\left(z\over1+z\right); \qquad y(z)=\tan^{-1} z; \qquad y(z) = {z\over 1+z^2}.
\end{equation}
These alternative choices seem to us to be not particularly well motivated in terms of the underlying physics, 
and from a mathematical perspective to also require significantly more subtle convergence analyses.

%----------------------------------------------------------------------------------------------------
\subsection{Hubble parameter $H(y)$}
%----------------------------------------------------------------------------------------------------

To see this in action, first note that in terms of this $y$-redshift one has
\begin{eqnarray}
\fl
1-y &=& 1 +H_0(t-t_0) - {q_0\over2!} [H_0(t-t_0)]^2 + {j_0\over3!} [H_0(t-t_0)]^3
+{s_0\over4!} [H_0(t-t_0)]^4
\nonumber
\\
\fl && \qquad\qquad \vphantom{1\over1}
+{c_0\over5!} [H_0(t-t_0)]^5+{p_0\over6!} [H_0(t-t_0)]^6+ O([t-t_0]^7). 
\end{eqnarray}
%%%%%%%%%%%%%%%%%%%%%
Reversion of this power series~\cite{reversion1,reversion2} yields:
\begin{eqnarray}
\fl t(y) &=& t_0 + {y\over H_0} \left\{-1+{1\over2!}q_0y -{1\over3!}\left(j_0-3q_0^2\right)y^2 
-
{1\over4!} \left(s_0 +10q_0j_0-15q_0^3\right)y^3 
\right.
\nonumber\\
\fl &&
\qquad \qquad \vphantom{1\over1}
+{1\over 5!}(c_0+15s_0q_0-10j_0^2+105j_0q_0^2-105q_0^4)y^4
\nonumber\\
\fl &&
\qquad\qquad
 - {1\over 6!}
 (p_0+21c_0q_0-35s_0j_0+210s_0q_0^2-280j_0^2 q_0+1260j_0q_0^3 -945q_0^4)y^5
 \nonumber\\
\fl &&
\left. 
\qquad\qquad\vphantom{1\over1}
 + O(y^6)  \right\}.
\end{eqnarray}
%%%%%%%%%%%%
Consequently for the Hubble parameter we now see
\begin{eqnarray}
\fl H(y) &=& H_0 \left\{1 + (1+q_0)y + \left[1+q_0 + {1\over2} (j_0-q_0^2)\right] y^2
\right.
\nonumber\\
\fl &&
\qquad\qquad  +
\left[1+q_0- {1\over3!}(s_0-3j_0 +4j_0q_0 +3q_0^2-3q_0^3 )\right] y^3 
\nonumber\\
\fl &&
\qquad\qquad  +
\left[1+q_0 + {1\over4!}  (c_0-4s_0+12j_0-  4j_0^2 + (7s_0-16j_0) q_0 
\right.
\nonumber\\
\fl &&
\left.
\qquad\qquad\qquad
+(25j_0-12) q_0^2 +12 q_0^3 -15 q_0^4)\vphantom{1\over1}\right] y^4 
\nonumber\\
\fl &&
\qquad\qquad  
+
\left[1+q_0- {1\over5!}((p_0-5c_0+20s_0-60j_0 +20j_0^2 -15j_0s_0)
\right.
\nonumber\\
\fl &&
\qquad\qquad\qquad
+
(11c_0-35s_0+80j_0-70j_0^2)q_0 +(60s_0-125j_0+60)q_0^2 
\nonumber\\
\fl &&
\qquad\qquad\qquad\left.\vphantom{1\over1}
+(210j_0-60)q_0^3 +75q_0^4-105 q_0^5  )\right] y^5 
\nonumber\\
\fl &&
\left. \qquad\qquad \vphantom{1\over1}
+ O(y^6)\right\}. 
\end{eqnarray}
Notice that this can be partially summed
\begin{eqnarray}
\fl H(y) &=& H_0 \left\{1 + (1+q_0){y\over1-y} + \left[ {1\over2} (j_0-q_0^2)\right] y^2
\right.
\nonumber\\
\fl &&
\qquad\qquad  -
\left[ {1\over3!}(s_0-3j_0 +4j_0q_0 +3q_0^2-3q_0^3 )\right] y^3 
\nonumber\\
\fl &&
\qquad\qquad  
+
\left[{1\over4!}(c_0-4s_0+12j_0-4j_0^2 + (7s_0-16j_0) q_0 
\right.
\nonumber\\
\fl &&
\left.
\qquad\qquad\qquad
+(25j_0-12) q_0^2 +12 q_0^3 -15 q_0^4)\vphantom{1\over1}\right] y^4 
\nonumber\\
\fl &&
\qquad\qquad  
-
\left[{1\over5!}((p_0-5c_0+20s_0-60j_0 +20j_0^2 -15j_0s_0)
\right.
\nonumber\\
\fl &&
\qquad\qquad\qquad
+ (11c_0-35s_0+80j_0-70j_0^2)q_0 +(60s_0-125j_0+60)q_0^2 
\nonumber\\
\fl &&
\qquad\qquad\qquad\left.\vphantom{1\over1}
+(210j_0-60)q_0^3 +75q_0^4-105 q_0^5  )\right] y^5 
\nonumber\\
\fl &&
\left. \qquad\qquad \vphantom{1\over1}
+ O(y^6)\right\}. 
\end{eqnarray}
The pole in this expression is at $y=1$ which corresponds to the big-bang singularity. 
Note, for instance, that at $z\approx 4$ we have $y \approx {4\over5} < 1$, so these $y$-expansions will be somewhat better behaved than the original $z$-expansions.

%----------------------------------------------------------------------------------------------------
\subsection{Deceleration, jerk, snap, crackle, and pop parameters in terms of $y$}
%----------------------------------------------------------------------------------------------------

Similarly the deceleration, jerk, snap, crackle, and pop parameters can now easily be expanded  in terms of the $y$-redshift.

\paragraph{Deceleration:}
\begin{eqnarray}
\fl q(y) &=& q_0 +(j_0-q_0-2 q_0^2) y
- {1\over2!}\left( s_0+2j_0+7 j_0q_0 - 4q_0^2 -8 q_0^3\right) y^2 
\nonumber\\
\fl &&
+{1\over3!}(c_0+3s_0-7j_0^2 +(11s_0+21j_0)q_0 +59 j_0 q_0^2 -24 q_0^3 -48q_0^4)y^3
\nonumber\\
\fl &&
-{1\over4!}(p_0+4c_0-25s_0j_0-28j_0^2 + (16c_0+44s_0-160j_0^2) q_0 + (125s_0+236j_0) q_0^2
\nonumber\\
\fl && \qquad\qquad
+605 j_0 q_0^3 -192 q_0^4 -384 q_0^5)y^4
+ O(y^5).
\end{eqnarray}

\paragraph{Jerk:}
\begin{eqnarray}
\fl j(y) &=& j_0 -(s_0 +2j_0+3 q_0j_0) y  
+ {1\over2!}(c_0+4s_0+2j_0-3j_0^2 +(7s_0+12j_0) q_0 
+15 j_0q_0^2) y^2
\nonumber\\
\fl &&
-{1\over 3!} (p_0+6c_0+6s_0-13j_0s_0-18j_0^2 
+ (12c_0+42s_0+18j_0- 48 j_0^2) q_0 
\nonumber\\
\fl &&
\qquad \qquad +(57s_0+90j_0) q_0^2 +105j_0 q_0^3) y^3
+O(y^4).
\end{eqnarray}

\paragraph{Snap:}
\begin{eqnarray}
\fl s(y) &=& s_0 - (c_0+3s_0+4s_0q_0) y +{1\over2!}\{p_0+6c_0+6s_0-4s_0j_0 +(9c_0+24s_0)q_0 + 24 s_0 q_0^2\}y^2 
\nonumber\\
\fl && + O(y^3).
\end{eqnarray}

\paragraph{Crackle:}
\begin{equation}
c(y) = c_0 - (p_0+4c_0+5c_0q_0)y+ O(y^2).
\end{equation}

\paragraph{Pop:}
\begin{equation}
p(y) = p_0 + O(y).
\end{equation}

%----------------------------------------------------------------------------------------------------
\subsection{Redshift drift in terms of $y$}
%----------------------------------------------------------------------------------------------------

\paragraph{First-order in $y$:} Note
\begin{equation}
\dot y = {\dot z\over(1+z)^2} = {(1+z)H_0-H(z)\over(1+z)^2} = (1-y)H_0 - (1-y)^2 H(y).
\end{equation}
This result is so far exact (within the context of FLRW cosmology).

Thence, inserting our cosmographic expansions, we explicitly have
\begin{eqnarray}
\fl \dot y &=& - H_0 y \left\{q_0 +{1\over2!}(j_0-2q_0-q_0^2)y 
- {1\over 3!} (s_0+3j_0+4j_0q_0-3q_0^2-3q_0^3) y^2\right.
\nonumber\\
\fl &&
+{1\over4!}(c_0+4s_0-4j_0^2 + (7s_0+16j_0) q_0  +25 j_0 q_0^2 -12 q_0^3 -15 q_0^5)y^3
\nonumber\\
\fl &&
-{1\over5!}( p_0+5c_0-15s_0j_0 -20j0^2 +(11c_0+35s_0-70j_0^2)q_0 
+(60 s_0+125 j_0)q_0^2 
\nonumber\\
\fl &&
 \qquad \left.\vphantom{1\over1} +210 j_0q_0^3 -75 q_0^4-105 q_0^5   )y^4+ O(y^5)\right\}.
\end{eqnarray}
This is our key cosmographic result for redshift drift in terms of the $y$-redshift. 

Note that at small redshift (where $y \approx z$) we have
\begin{equation}
\dot y = - y \,q_0\, H_0 + O(y^2).
\end{equation}

\paragraph{Second-order in $y$:} Similarly
\begin{equation}
\fl \ddot y = {d\over dt} \left( \dot z\over(1+z)^2\right)
 = {\ddot z\over(1+z)^2} -  {2(\dot z)^2 \over(1+z)^3}
 = (1-y)^2 \ddot z - 2 (1-y)^3 (\dot z)^2.
\end{equation}
Thence, substituting $z\to {y\over1-y}$ in our previous expressions for $\ddot z$ and $\dot z$ one has
\begin{eqnarray}
\fl \ddot y &=& y H_0^2 \left\{j_0 -{1\over2!}(s_0+3j_0+j_0q_0-3q_0^2) y \right.
\nonumber\\
\fl && +  {1\over3!} (c_0+5s_0+3j_0-j_0^2 +(3s_0+14j_0)q_0 +(3j_0-9)q_0^2 -9q_0^3) y^2 
\nonumber\\
\fl &&
- {1\over 4!} (p_0+7c_0+8s_0-5s_0j_0-16j_0^2 
+(6c_0+33s_0+44j_0-10j_0^2)q_0 
\nonumber\\
\fl &&\left. \vphantom{1\over1}\qquad
+(15s_0+87j_0)q_0^2 +(15j_0-36) q_0^3 -45q_0^4) y^3  + O(y^4)  \right\}.
\end{eqnarray}
Note that at small redshift (where $y \approx z$) we have
\begin{equation}
\ddot y =  y\, j_0\, H_0^2 + O(y^2).
\end{equation}

\paragraph{Third-order in $y$:}
The pattern should now be clear. For the next derivative
\begin{equation}
\dddot y = {d\over dt} \left( {\ddot z\over(1+z)^2} -  {2(\dot z)^2 \over(1+z)^3}\right)
 = {\dddot z\over(1+z)^2} -  {6\ddot z \dot z \over(1+z)^3} 
 + {6 (\dot z)^3 \over (1+z)^4}.
\end{equation}
That is
\begin{equation}
\dddot y = {(1-y)^2 \dddot z} -  {6 (1-y)^3 \ddot z \dot z} 
 + {6(1-y)^4 (\dot z)^3}.
\end{equation}

But, given that we already know that $\dddot z$, $\ddot z$, and $\dot z$, we simply substitute $z\to {y\over1-y}$ which implies
\begin{eqnarray}
\fl \dddot y &=& y H_0^3 \left\{ s_0 - {1\over2!}(c_0+4s_0+s_0q_0) y
\right.
\nonumber\\
\fl &&
+{1\over3!}(p_0+7c_0+8s_0-s_0j_0-j_0^2 +(3c_0+6s_0-j_0)q_0+3 s_0q_0^2+7779q_0^3)y^2
\nonumber\\
\fl &&
\left.\vphantom{1\over1}
 + O(y^3)
\right\}.
\end{eqnarray}
At small redshift
\begin{equation}
\dddot y = y \,s_0 \, H_0^3   + O(y^2).
\end{equation}

\paragraph{Fourth-order in $y$:}
At fourth order
\begin{equation}
\ddddot y = {d\over dt} \left(  {\dddot z\over(1+z)^2} -  {6\ddot z \dot z \over(1+z)^3} 
 + {6 (\dot z)^3 \over (1+z)^4} \right).
\end{equation}
Thence
\begin{equation}
\ddddot y = 
 {\ddddot z\over(1+z)^2} -  
 {8 \dddot z \dot z +6(\ddot z)^2 \over(1+z)^3} 
 +{36\ddot z (\dot z)^2\over(1+z)^4}
 - {24 (\dot z)^4 \over (1+z)^5}.
\end{equation}
That is 
\begin{equation}
\fl\qquad \ddddot y = 
 (1-y)^2 \ddddot z -  
 8 (1-y)^3 [\dddot z \dot z +6(\ddot z)^2]
 +36(1-y)^4\ddot z (\dot z)^2
 - 24 (1-y)^5(\dot z)^4.
\end{equation}
But we have already evaluated each of these ingredients. So we can simply substitute $z\to {y\over1-y}$, which now implies
\begin{eqnarray}
\ddddot y &=& y H_0^4 \left\{ c_0 - {1\over2!}(p_0+5c_0+c_0q_0 + 3s_0q_0-14j_0^2) y
 + O(y^2)
\right\}.
\end{eqnarray}
At small redshift
\begin{equation}
\ddddot y = y \,c_0 \, H_0^4   + O(y^2).
\end{equation}

\paragraph{Fifth-order in $y$:}
At fifth-order in $y$ we use a minor variant of the result for fifth-order in $z$. 
We simplify the argument by considering
\begin{equation}
y^{(5)} = {d\over dt} \left\{ \ddddot y\right\} 
= {d\over dt} \left\{  y \, c_0 \, H_0^4 + O(y^2)\right\}
= {d\over dt} \left\{  y \, c_0 \, H_0^4\right\} +O(y^2).
\end{equation}
Here we have used the fact that $\dot y = O(y)$.
Then
\begin{equation}
y^{(5)} 
= \left\{  \dot y \, c_0 \, H_0^4+ y \dot c_0 H_0^4 + 4 y c_0 H_0^3 \dot H_0\right\} +O(y^2).
\end{equation}
But $\dot y = -y q_0 H_0+O(y^2)$ and $\dot H_0 = -(1+q_0)H_0^2$ while
\begin{equation}
\dot c_0 = (p_0 + (4+5 q_0) c_0) H_0.
\end{equation}
Combining the above
\begin{equation}
y^{(5)} = y \, p_0 \, H_0^5 + O(z^2).
\end{equation}
We shall now extend this to a general $n^{th}$-order result.

\paragraph{$n^{th}$-order in $y$:}
Finally we point out that to lowest order in $y$, a minor variant of the argument used for $z$ yields
\begin{equation}
y^{(n)} = y \;k_{n+1} \; H_0^n + O(y^2); \qquad \forall n \geq 1. 
\end{equation}
This completes our cosmographic analysis for redshift drift in terms of the $y$-redshift.

%----------------------------------------------------------------------------------------------------
\section{Discussion and Conclusions}
%----------------------------------------------------------------------------------------------------
\label{S:Conclusions}
%----------------------------------------------------------------------------------------------------

What we have seen above is that the main central features of the redshift drift can be dealt with cosmographically, using only the symmetries of the FLRW spacetime. 
We have explicitly included present-epoch values of the jerk, snap, crackle, and pop parameters in the redshift-dependent cosmographic expansion
for all of the Hubble, deceleration, jerk, snap, crackle, and pop parameters, and for the redshift drift and its derivatives $\dot z$, $\ddot z$, $\dddot z$, $\ddddot z$, and $z^{(5)}$. One could in principle go to even higher order, the relevant formulae just become messy and tedious. 
We have also derived a quite general result for $z^{(n)}$ at lowest order in $z$.

Since in applications one often wants to work at large redshift ($z>1$) we have shown how to ameliorate problematic convergence issues by rephrasing the discussion in terms of a modified notion of redshift, the $y$-redshift $y={z\over1+z}$. 
All of our cosmographic expansions have also been expressed in terms of the $y$-redshift.

Of course very much more could be said by introducing \emph{cosmo-dynamics},  (that is, invoking the Einstein equations, or more specifically the Friedmann equations).
This would first allow us to build specific explicit analytic models for $a(t)$ in ideal FLRW spacetimes, thereby largely side-stepping the cosmographic expansion. Subsequently we also plan to consider deviations from ideal FLRW universes, (peculiar motions, density fluctuations, \emph{etcetera}), but we shall leave all such considerations for future work.

%------------------------------------------------
\subsection*{Acknowledgments}
FSNL and JPM acknowledge funding from the Funda\c c\~{a}o para a Ci\^encia e a Tecnologia (FCT, Portugal) research projects No. UID/FIS/04434/2019 and No.~PTDC/FIS-OUT/29048/2017. FSNL also thanks support from the FCT Scientific Employment Stimulus contract with reference CEECIND/04057/2017.
MV was supported by the Marsden Fund, via a grant administered by the Royal Society of New Zealand.

%\clearpage
%=======================================================
\section*{References}
%========================================================

%========================================================
\end{document}